\documentclass{eas} 
\usepackage{graphicx}
\newcommand{\be}{\begin{equation}}
\newcommand{\ee}{\end{equation}}
\newcommand{\gs}{\;\raisebox{-.8ex}{$\buildrel{\textstyle>}\over\sim$}\;}
\newcommand{\ls}{\; \raisebox{-.8ex}{$\buildrel{\textstyle<}\over\sim$}\;}

\newcommand{\apj}{{\it ApJ, }}

\newcommand{\icar}{{\it Icarus, }}
\newcommand{\mnr}{{\it MNRAS, }}

\newcommand{\ana}{{\it A\&A, }}

\begin{document}

\title{Conditions for the occurrence of  mean-motion resonances 
in  a low mass planetary system} 
\author{J.C.B. Papaloizou}\address{Department of Applied Mathematics and Theoretical Physics, 
Centre for Mathematical Sciences, Wilberforce Road, Cambridge CB3 0WA, UK}
\author{E. Szuszkiewicz}\address{  CASA* and Institute of Physics, 
University of Szczecin, Wielkopolska 15, 70-451 Szczecin, Poland}
%
\runningtitle{Papaloizou \& Szuszkiewicz: MMRs in a low mass planetary system}
\begin{abstract}
The dynamical interactions that occur in newly formed planetary systems
may reflect the conditions occurring in the protoplanetary disk
out of which they formed.
With this in mind, we explore the attainment and maintenance of orbital resonances
by migrating planets in the terrestrial mass range. Migration time scales varying between
$\sim 10^6 $~yr and $\sim 10^3 $~yr are considered. In the former case, for which the
migration time is comparable to the lifetime of the  protoplanetary gas disk,
a 2:1 resonance may be formed. In the latter, relatively rapid migration
regime commensurabilities of high degree such as 8:7 or 11:10 may be formed.
However, in any one large-scale migration  several different
commensurabilities may be formed sequentially, each being associated with 
significant orbital evolution. 
We also use a simple analytic theory to develop conditions for
first order commensurabilities to be formed. These depend on the
degree of the commensurability, the imposed migration and circularization rates,
and the planet mass ratios. These conditions are found to be consistent with the
results of our simulations.
 
\end{abstract}
\maketitle
\section{Introduction}
The increasing number of extrasolar multi-planet systems
 have stimulated studies  of their  origin, 
evolution and  stability.  An important feature
 in  planetary system evolution is the occurrence of 
mean-motion resonances, which may relate to conditions
at the time of or just after the process of  formation.
There are  some well known examples of systems 
 such as   Gliese 876 (\cite{mbfv00}),
 HD 82943 (\cite{munp00}) and 55 Cancri (\cite{mmfq00})
 involving  planets with
masses in the Jovian  range. More recently
a system of three superearths orbiting HD40307  has been announced for which
the period ratios are not strictly commensurable but are close to
 a 2:1 commensurability (\cite{mulp08}). 

It is therefore  important to establish the main  features of the
evolution of low mass planets embedded in  a gaseous disc and  in
particular  to determine the types of resonant  configurations that
might arise when a pair of such planets evolves together.
The disc-planet interaction naturally produces orbital migration
through the action of tidal torques (\cite{gt80}, \cite{lpap86})
which in turn  may lead to an orbital resonance in a many planet system
(eg. \cite{spn01}, \cite{lp02}).
This is because it is expected  that
two planets with different masses will migrate at different rates.
This has the consequence that their period ratio will evolve
with time. 
In the situation where the migration is such that the orbits
converge,   they tend to enter and  become  locked in a mean-motion resonance
(eg. \cite{np02}, \cite{kpb04}) and then subsequently migrate together.

\noindent In the simplest case of nearly
circular and coplanar orbits the resonances that are formed  are
the first-order resonances which occur at locations
where the ratio of the two orbital periods 
can be expressed as $(p+1) / p$,  with $p$
being an integer.

Papaloizou \& Szuszkiewicz (2005) performed a recent analytic and numerical study
of the formation of first order commensurabilities
by a system of two planets in the earth mass range migrating in a laminar disc.
Here we extend these studies using $N$-body simulations to a larger range
of migration rates and commensurabilities and compare the numerical work 
to  analytically derived conditions for particular commensurabilities to form.
We begin by deriving the analytic criteria and go on to present
the results of simulations to which they are reconciled.
We consider migration time scales ranging between $\sim 10^3 $~yr and $\sim 10^6 $~yr.
For the longest time scales,  2:1 commensurabilities may be set
up while for the shortest, commensurabilities as high as 11:10 have been found.
Finally we summarize our results.

\section{Basic Equations}
The equations of motion for a system of $N$ planets $(i \ \ = \ \ 1, 2, 3,.... N)$
moving in a fixed plane, about a dominant central mass,
under a general Hamiltonian $H$ may be written in the form
(see e.g. \cite{pa03}, \cite{ps05}) 
\begin{eqnarray}
\dot E_i &=& -n_i\frac{\partial H}{\partial \lambda_i}\label{eqnmo1}\\
\dot L_i &=& -\left(\frac{\partial H}{\partial \lambda_i}+\frac{\partial H}{\partial \varpi_i}\right)\\
\dot \lambda_i &=& \frac{\partial H}{\partial L_i} + n_i \frac{\partial H}{\partial E_i}\\
\dot \varpi_i &=& \frac{\partial H}{\partial L_i}.\label{eqnmo4}
\end{eqnarray} 
Here the orbital  angular momentum of  planet  $i$ which has
reduced mass
$m_i = m_{i0}M/(M+m_{i0}),$ with $m_{i0}$ being the actual mass,  is $L_i$ and the
orbital  energy is $E_i.$
For motion around a central point  mass $M$ we have  
\begin{eqnarray}
     L_i &=&  m_{i}\sqrt{GM_{i}a_i(1-e_i^2)} \hspace{0.5cm}  {\rm and} \\
     E_i &=& -{{GM_{i}m_{i}}\over{2a_i}}.
\end{eqnarray}
Here $M_{i} = M+m_{i0},$  $a_i$ denotes the semi-major axis and $e_i$  the eccentricity
of planet $i.$

\noindent The  mean longitude of planet $i$ is $\lambda_i = n_i (t-t_{0i}) +\varpi_i ,$ 
 where $n_i  = \sqrt{GM_{i}/a_i^3}$ is its mean motion, with
$t_{0i}$ denoting its time of periastron passage and $\varpi_i$ the longitude of periastron.
  
\noindent In this paper we consider $N=2$ and arrange the two planets
such that $i=1$ denotes the outer planet and $i=2$ denotes the inner planet.
We remark that the above formalism does not incorporate disk tides. However,
forces resulting from these can be added in separately (see below).

\subsection{Coordinate system}
We adopt Jacobi coordinates (\cite{si75}, \cite{ps05}) for which the radius vector
 ${\bf r}_2,$ 
is measured from  $M$
and that of the  outer planet, ${\bf r }_1,$ is referred
to the  centre of mass of  $M$ and  the inner planet $i = 2.$
The required Hamiltonian  correct to second order in the planetary masses  is given by 
\begin{eqnarray} H & = &  {1\over 2} ( m_1 | \dot {\bf r}_1|^2 +m_2| \dot {\bf r}_2|^2)
- {GM_{1}m_1\over  | {\bf r}_1|} - {GM_{2}m_2\over  | {\bf r}_2|} \nonumber \\
& - &{Gm_{1}m_2\over  | {\bf r}_{12}|}
 +  {Gm_{1}m_2 {\bf r}_1\cdot {\bf r}_2
\over  | {\bf r}_{1}|^3}.
\end{eqnarray}
Here $M_{1}=M+m_1, M_{2}= M + m_2 $ and
$ {\bf r}_{12}= {\bf r}_{2}- {\bf r}_{1}.$ 
 
\noindent The Hamiltonian may quite generally
 be expanded in a Fourier series
involving linear combinations of the three angular differences
$\lambda_i - \varpi_i, i=1,2$ and $\varpi_1 -\varpi_2.$

Near a first order  $p+1 : p $ resonance, we expect that both
$\phi_1 = (p+1)\lambda_1-p\lambda_2-\varpi_2, $ and
$\phi_2 = (p+1)\lambda_1-p\lambda_2-\varpi_1,$
will be slowly varying.
 Following  standard practice
 (see e.g. \cite{pa03} and  \cite{ps05})
only terms in the Fourier expansion involving  linear 
combinations of $\phi_1$ and $\phi_2$
as argument are  retained because 
only  these are expected to lead to large long-term perturbations.
In general there are an infinite number of such terms that need to 
be considered. However, in the limit of small eccentricities
only terms that are first order in the eccentricities need to be retained.
This approximation is valid when the circularization times are small enough
to ensure that the eccentricities remain small. This situation is realized
for realistic  examples of  low mass protoplanets migrating in protoplanetary discs
(\cite{ps05}).

\noindent Following the procedure described above 
and expanding to first order in the eccentricities, the  Hamiltonian   may  be written
in the form $H=E_1+E_2+ H _{12},$ where
\begin{equation} H _{12}= -\frac{Gm_1m_2}{a_1}\left(  e_1 C_1\cos (\phi_2)
- e_2C_2\cos (\phi_1) \right), \label{Hamil} \end{equation}
with
\begin{equation} C _{1}={1 \over 2}\left(   x{d(b^{(p)}_{1/2}(x))\over dx} +(2p+1)b^{(p)}_{1/2}(x)
-4x\delta^{p}_1 \right) \ \ {\rm and} \ \ \label{Hamil1} \end{equation}
\begin{equation} C _{2}={1 \over 2}\left(   x{d(b^{(p+1)}_{1/2}(x))\over dx} +2(p+1)b^{(p+1)}_{1/2}(x)  \right) .\label{Hamil2} \end{equation}
Here $b^{(p)}_{1/2}(x)$ denotes the usual Laplace coefficient
(e.g. Brouwer \& Clemence 1961)
with the argument $x = a_2/a_1$ and $\delta^{p}_1$ denotes
the Kronecker delta. 
We have also 
replaced $M_{i}$ by $M.$
\subsection{Behaviour near a resonance with disk tides incorporated}
The governing equations for motion near to the $p+1:p $
resonance are to lowest order in the eccentricities
\begin{eqnarray}
\!\!\!\!\!\!\!\!\!\!
\frac{1}{2}\frac{d e_1^2}{dt} &\!\!\!\!\!\!\!\!\!\!\;\;= &\;\; 
\!\!\!\!\!\!\!\!\!\!
-\frac{Gm_2 C_1}{a_1\sqrt{GMa_1}}e_1\sin\phi_2 - \left[\frac{e_1^2}{t_{c1}}\right] \label{eqne1}\\
\!\!\!\!\!\!\!\!\!\!
\frac{1}{2}\frac{d e_2^2}{dt} &\!\!\!\!\!\!\!\!\!\!\;\; =&\;\;
\!\!\!\!\!\!\!\!\!\! 
-\frac{Gm_1 C_2}{a_1\sqrt{GMa_2}}e_2\sin\phi_1-\left[\frac{e_2^2}{t_{c2}}\right]
\label{eqne2}\\
\!\!\!\!\!\!\!\!\!\!
\dot n_1 &\!\!\!\!\!\!\!\!\!\!\;\; =&\;\;
\!\!\!\!\!\!\!\!\!\! 
\frac{3n_1(p+1)Gm_2}{a_1\sqrt{GMa_1}}\left(C_1e_1\sin\phi_2-C_2e_2\sin\phi_1 \right)
+\left[\frac{n_1}{t_{mig1}}+\frac{3n_1e_1^2}{t_{c1}}\right] \label{eqntid0}\\
\!\!\!\!\!\!\!\!\!\! 
\dot n_2 &\!\!\!\!\!\!\!\!\!\!\;\;= &\;\;
\!\!\!\!\!\!\!\!\!\!
-\frac{3n_2pGm_1}{a_1\sqrt{GMa_2}}\left(C_1e_1\sin\phi_2-C_2e_2\sin\phi_1 \right) 
+\left[\frac{n_2}{t_{mig2}}+\frac{3n_2e_2^2}{t_{c2}}\right].\label{eqntid} \\
\nonumber
\end{eqnarray} 
Here the terms on the right-hand sides enclosed in square brackets give the contributions arising from disk tides. These have been discussed elsewhere  (see e.g. \cite{ps05}).
The circularization and migration times for planet $i$ are $t_{ci}$ and $t_{migi}$ 
respectively. Note that the migration time is defined here as the time for the mean motion
to change by a factor of $e$ as a result of disk torques. The terms $\propto e_i^2/t_{ci}$
in (\ref{eqntid0}) and (\ref{eqntid}) account for the orbital energy dissipation occurring as a result of circularization
at the lowest order in $e_i$ which it appears.

The other terms on the right-hand sides of  (\ref{eqne1})-(\ref{eqntid}) are derived from 
the Hamiltonian system (\ref{eqnmo1})-(\ref{eqnmo4}) using the Hamiltonian (\ref{Hamil}) 

\subsection{Migration maintaining resonance}
When the two planets migrate together maintaining resonance,
we expect and find solutions for which $e_1$ and $e_2$ are either actually nearly constant
or more generally very nearly constant in a time average sense
while the ratio $n_1/n_2$ is maintained very close to $p/(p+1).$ 
When considering time averages, we here consider the average
to be taken over a time long compared to the characteristic orbital period but short enough
that the semi-major axes  and tidal time scales may be considered constant.
  Using angle brackets
to denote such a time average 
and setting $\langle e_1 \dot e_1\rangle  =\langle e_2 \dot e_2\rangle =0$
in (\ref{eqne1}) and (\ref{eqne2}) and taking the result
of  dividing the time averaged (\ref{eqntid0}) by the time averaged  (\ref{eqntid})
  gives three equations from which $\langle e_i\sin(\phi_i)\rangle , \ \ i =1,2 \ \ $ may be determined together
  with a constraint on the mean square  eccentricities and migration times that does not depend
  on $\phi_i.$ Proceeding in this way
  we obtain
\begin{eqnarray}
\langle e_1 \sin\phi_2\rangle &=& - \frac{e_1^2a_1\sqrt{GMa_1}}{Gm_2 C_1t_{c1}}\label{eqne3}\\
 \langle e_2 \sin\phi_1\rangle &=& - \frac{e_2^2a_1\sqrt{GMa_2}}{Gm_1 C_2t_{c2}},
\label{eqne4}
\end{eqnarray}  
together with the constraint
\begin{equation}
 {e_1^2\over t_{c1}} + {e_2^2\over t_{c2}}{m_2n_1a_1
\over m_1n_2a_2} 
 -\left({e_1^2\over t_{c1}}-{e_2^2\over t_{c2}}\right)f
=
\left({1\over t_{mig1}}-{1\over t_{mig2}}\right){f\over 3},  
\label{ejcons}
\end{equation}
where  $ f =m_2a_1/((p+1)(m_2a_1+m_1a_2))$
and for ease of notation we have suppressed the angle brackets 
enclosing $e_i^2$ which is now implicitly assumed to be a time averaged value.  
We remark that (\ref{ejcons})   has already 
obtained from general considerations (\cite{ps05}).
However, (\ref{eqne3}) and (\ref{eqne4})  have not.
\subsection{ A condition for a p+1:p  commensurability to be maintained}
We may use (\ref{eqne3}) and (\ref{eqne4}) to express $e_i^2$  in terms of $\langle e_i \sin\phi_i\rangle $
in the constraint (\ref{ejcons}). Using Schwartz's inequality and 
applying the condition $\langle \sin^2\phi_i\rangle  \le 1$
results in an inequality that must be satisfied if the  resonance is to be maintained.
This inequality can be written in the form
\begin{equation}
 {p^2 n_2^2 m_2\over(p+1)^2 M} 
 \left({(1-f)m_2C_1^2t_{c1}\over M}+{m_1a_1^2C_2^2t_{c2}\over Ma_2^2}\right)
\ge
\left({1\over t_{mig1}}-{1\over t_{mig2}}\right){f\over 3}.
\label{ehcons}
\end{equation}
For the simple example where $m_1 \gg m_2$ is in a prescribed
slowly shrinking circular orbit
and controls the migration ($t_{mig2} \gg t_{mig1}$) the relation (\ref{ehcons})
simplifies to the form
\begin{equation}
 { m_1^2\over M^2} \ge 
 \left({a_2\over 3p a_1 n_1n_2 t_{mig1} t_{c2} C_2^2}\right).
\label{eicons}
\end{equation}
We further remark that when it is satisfied, the  maximum  $\langle \sin^2\phi_i\rangle$ 
has to exceed the ratio of the right-hand side to the  left-hand side
of the inequality (\ref{ehcons}).
\subsection{Resonance overlap}
Because it is found that both $C_1$ and $C_2$ increase with $p,$ while $f$ decreases with $p,$
the inequality (\ref{ehcons}) indicates that for given planet masses
 the maintenance of resonances
with larger values of $p$ is favoured. However,
the maintenance  of resonances with large $p$ may be prevented by resonance overlap
and the onset of chaos.
\noindent   
Resonance overlap  occurs when the difference of the semi-major 
axes of the two planets is below a limit  that, 
in the case of two equal mass planets, has a half-width given by \cite{gl93} as
\begin{equation}
{\Delta a\over a} \sim {2\over 3p} \approx 2 
\left({m_{\mathrm{planet}} \over M_*}\right)^{2/7},
\label{chao}
\end{equation}
with $a$ and $m_{\mathrm{planet}}$ being the mass and  semi-major axis 
of either planet respectively.
Thus for a system consisting of two equal four-Earth-mass planets 
orbiting a central solar mass we expect resonance overlap for 
$p \gs 8.$ Conversely we might expect isolated resonances in 
which systems of planets can be locked and migrate together 
if  $p  \ls 8.$ But note that the existence of eccentricity damping 
may allow for somewhat larger values of $p$ in some cases.
In this context the inequality (\ref{ehcons}) also suggests that resonances
may be more easily maintained for lower  circularization rates.
However, this  may be nullified for large $p$ by the tendency
for larger eccentricities to lead to greater instability.
Note also that higher order commensurabilities may also be generated
in such cases  and these are not covered by the theory described above.
\begin{table*}
\caption{ \label{table1} Details of the simulations discussed in this paper.
Column 1 gives the value of $f_{\mathrm{tid}}.$  Column 2 gives
the run time in years. Column 3 lists the commensurabilities for which sustained
trapping and joint migration was noted. The occurrence of such episodes
was characterized by  eccentricity boosts.
Column 4 firstly indicates the first order commensurability of lowest degree
estimated to be possible by use of  the inequality (\ref{ehcons})
and secondly the same estimate assuming the maximum allowed
$\langle \sin^2\phi_i\rangle \le  0.1.$ All the simulation results are  consistent
with the inequality (\ref{ehcons}).}

\centering
\begin{tabular}{|r|r|r|c|}
\hline
              &             &                   &             \\
 $f_{\mathrm{tid}}$    &Run time yr. &Commensurabilities &Lowest degree expected \\
              &             &                   &               \\
\hline
1   &1278400  & 2:1 3:2  & 2:1, 2:1  \\
2   &638400  &2:1 3:2   &2:1, 2:1\\
4      &638400   &2:1 3:2    &2:1, 2:1 \\
16    &492800      &2:1, 3:2, 4:3    &2:1, 2:1\\
32      &345600      &2:1, 3:2, 4:3       &2:1, 2:1\\
64    &123200   &3:2, 4:3, 5:4       &2:1,2:1 \\   
128       &57000       &3:2, 4:3, 5:4, 6:5       & 2:1, 3:2 \\
256   &31150   &  4:3, 5:4, 6:5       &2:1, 3:2\\
512     &15550      &4:3, 5:4, 6:5       &3:2, 5:4 \\
1024     &7875
      &5:4, 6:5, 7:6   &4:3,7:6\\
2048    &3906   &6:5, 7:6, 8:7   & 5:4, 10:9\\
4096    &1959   &11:10       & 8:7, $p > 12$\\
\hline
\end{tabular}
\end{table*}

\begin{figure*}[htp]
\centering
\includegraphics[width=2.5in,height=4.75in,angle=270]{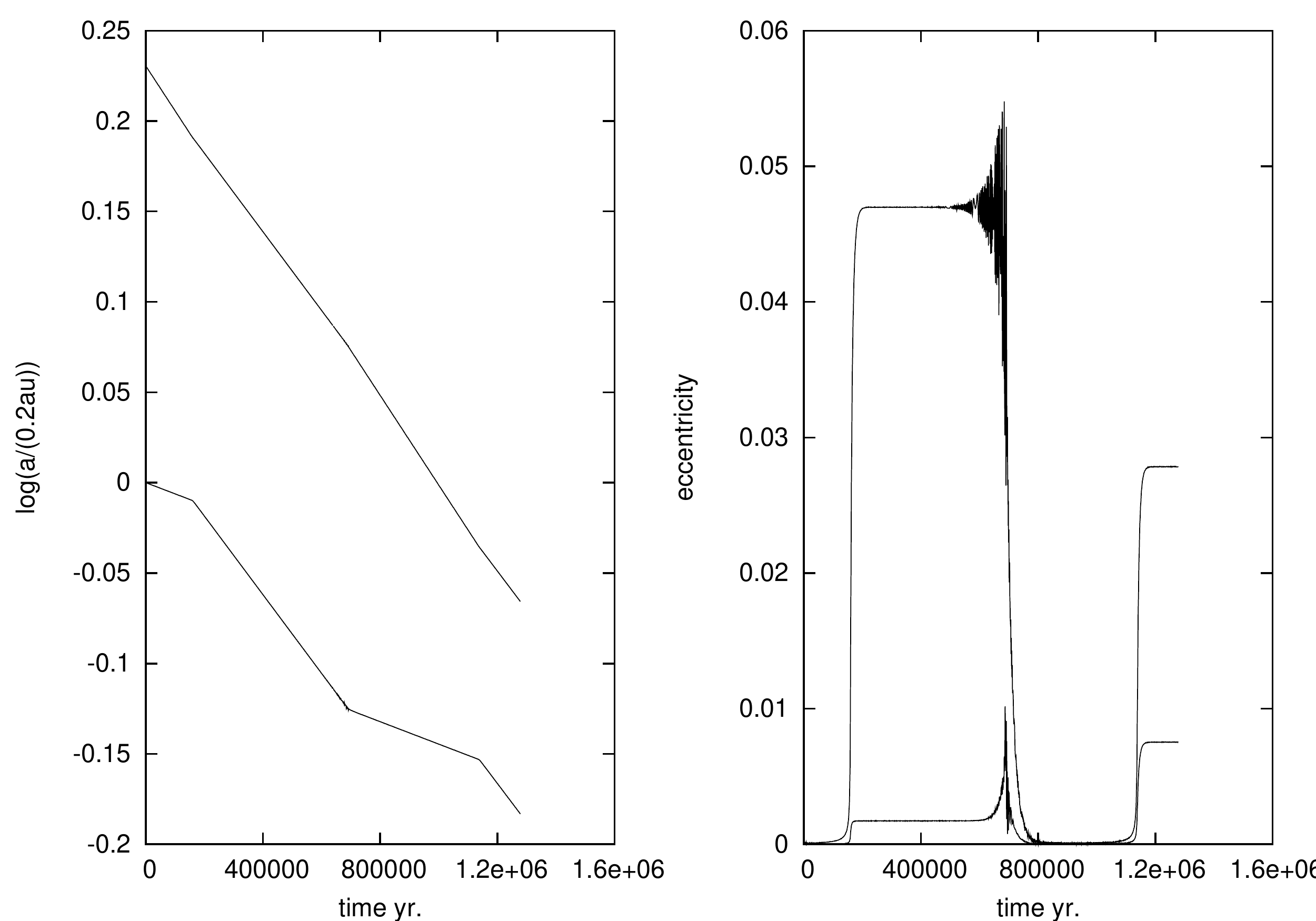}
\caption{The evolution of the semi-major axes (left panel) and
eccentricities (right panel) of the two planets are shown as functions
of time for $f_{\mathrm{tid}}=1.$
 The planets migrate inwards while locked in commensurabilities.
 The early episode of
high eccentricity corresponds to a 2:1 commensurability
while the final one corresponds to a 3:2 commensurability.
The inner planet has the larger eccentricity. }\label{fig:1}
\end{figure*}

\begin{figure*}[htp]
\centering
\includegraphics[width=2.5in,height=4.75in,angle=270]{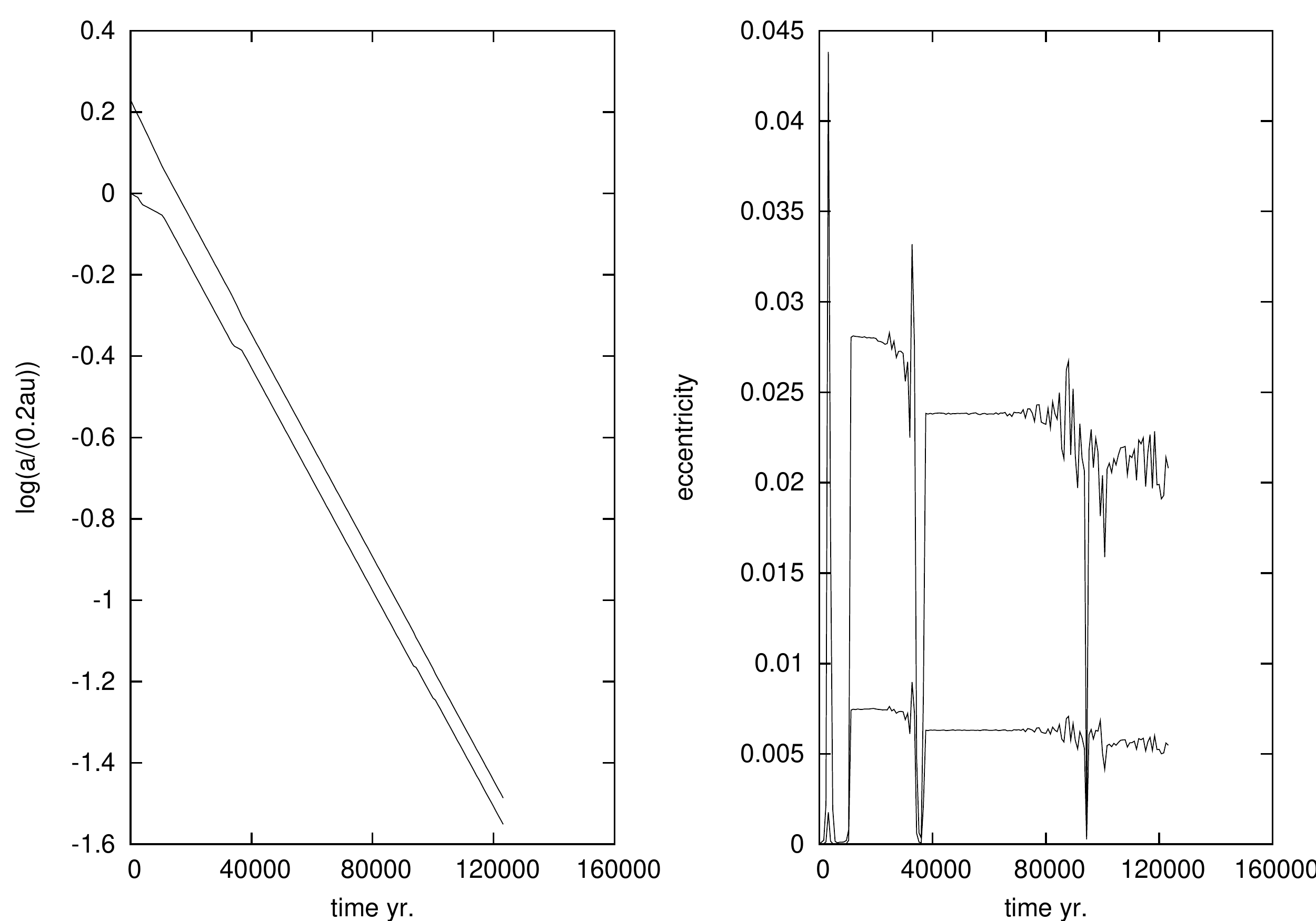}
\caption{As in Fig. \ref{fig:1} but for $f_{\mathrm{tid}}=64$.
In this case the planets fail to lock into the 2:1 commensurability,
showing only an eccentricity spike when the system passes through 
this commensurability.
Subsequently the configuration becomes sequentially locked
in the 3:2, 4:3 and 5:4 commensurabilities.}\label{fig:64}
\end{figure*}

\begin{figure*}[htp]
\centering
\includegraphics[width=2.5in,height=4.75in,angle=270]{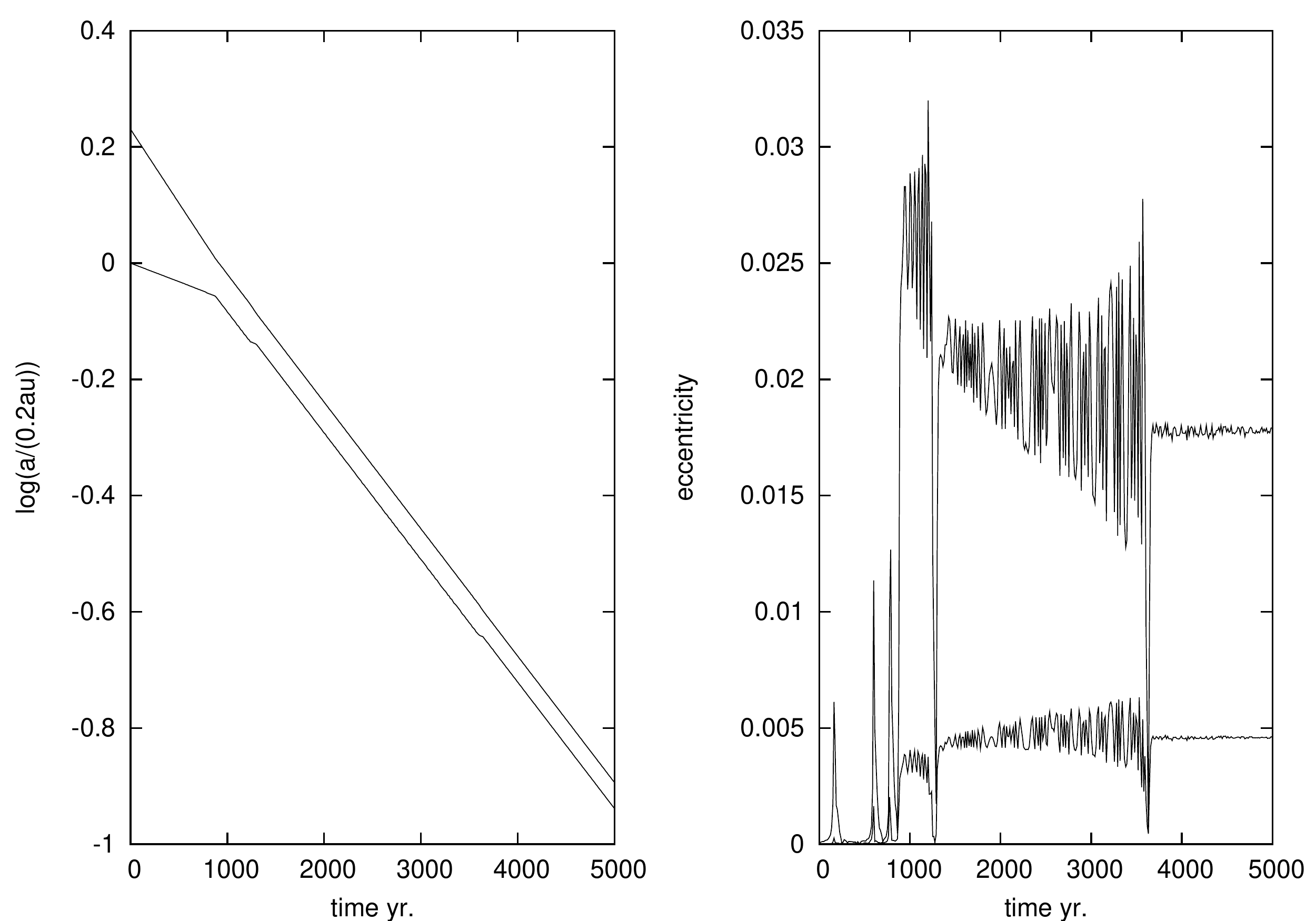}
\caption{As in Fig. \ref{fig:1} but for $f_{\mathrm{tid}}=1024.$
 In this case the planets fail to lock into the 2:1, 3:2 and 4:3 commensurabilities
showing only  eccentricity spikes when the system passes through these 
commensurabilities.
Subsequently the configuration becomes sequentially locked
in the 5:4, 6:5 and 7:6 commensurabilities.}\label{fig:1024}
\end{figure*}


\begin{figure*}[htp]
\centering
\includegraphics[width=2.5in,height=4.75in,angle=270]{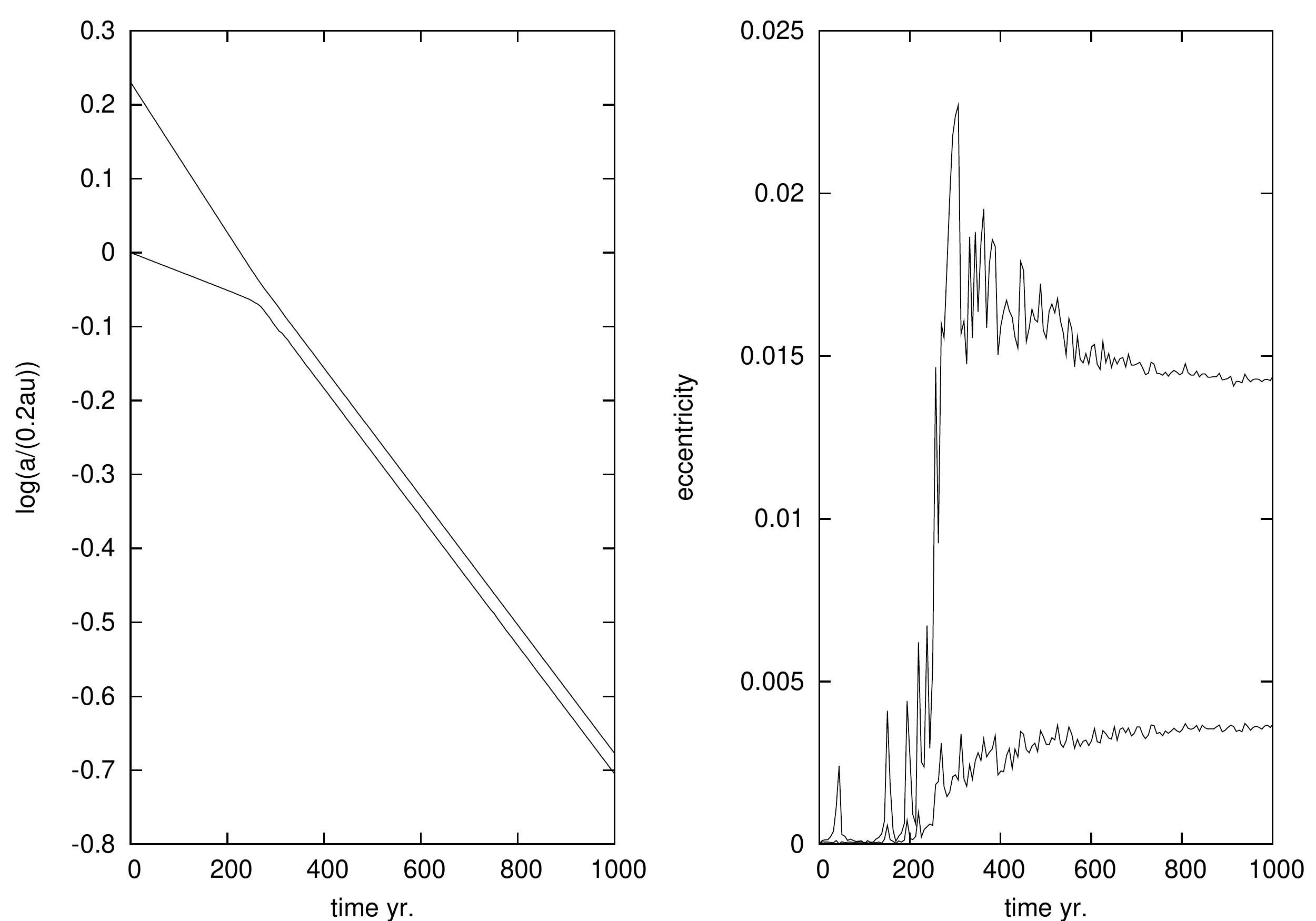}
\caption{As in Fig. \ref{fig:1} but for $f_{\mathrm{tid}}=4096.$
 In this case the planets fail to lock into any commensurability
 until $p=10$, corresponding to the 11:10 commensurability.
Many  eccentricity spikes occur as the system passes through
lower degree commensurabilities.}\label{fig:4096}
\end{figure*}

\section{Numerical  simulations}
In order to check the applicability of the inequality
(\ref{ehcons}) we have performed numerical simulations of a pair of migrating planets
using an  $N$-body code. The approach is 
the same as that used by many authors 
(e.g. Snellgrove, Papaloizou \& Nelson 2001;
\cite{np02}; 
\cite{lp02};
Kley, Peitz \& Bryden 2004).
 The reader is referred to these papers for details.
In particular, 
orbital migration and eccentricity damping
that are presumed to result from interaction with the protoplanetary
disk are
modelled 
 through the addition of appropriate
acceleration terms to the equations of motion. We considered planets with masses
$m_1 = 4M_{\oplus},$ and $m_2 = M_{\oplus}$ for a wide range
of imposed migration and circularization rates. The planets were started in circular orbits
with $m_1$ being at $0.34au$ and $m_2$ at $0.2au$  (note that these
results can be scaled to apply to other initial radii by appropriate
renormalization of space and time scales). Thus they begin just outside of the 2:1 resonance
before commencing convergent migration.
Low mass planets of the type we consider undergo type I migration (see \cite{pt06} and
references therein).
 On account of dependence 
on the equation of state and non linear effects,
the precise rate   that should be employed  is uncertain
even in a laminar disk as adopted here (e.g. \cite{pp08}).
To survey a range of possibilities, the migration  and circularization rates 
 we adopted  were given by
 \be t_{migi}=\frac{1.4 \times 10^7}{3f_{\mathrm{tid}}}\frac{M_{\oplus}}{m_i}
     \ \ yr\ee
and
  \be t_{ci}=\frac{2.5\times 10^4}{f_{\mathrm{tid}}} \frac{M_{\oplus}}{m_i} \ \ yr, \ee
respectively where $f_{\mathrm{tid}}$ is a scaling factor. We have adopted $1 \le f_{\mathrm{tid}} \le 4096,$
giving migration time scales  that range between $\sim 10^6 $~yr and $\sim 10^3 $~yr.
In this way timescales that are both comparable to the gas disk lifetime, and significantly
shorter are considered. The ratio  $ t_{ci}/t_{migi}$ has been chosen to be comparable
to the disk aspect ratio squared for an aspect ratio $\sim 0.05$ as expected
from theoretical work (see e.g. \cite{pl00})

\noindent 
We summarize the parameters and outcomes of  the  
simulations in Table \ref{table1}.
    In particular the  commensurabilities for which sustained
trapping and joint migration  were observed are noted. 
As expected, it is found that lower degree commensurabilities are formed for smaller
$f_{\mathrm{tid}}$ and then become unsustainable if $f_{\mathrm{tid}}$ is increased beyond
a critical level. In general each simulation is found to involve
the setting up of a sequence of commensurabilities  each of which  is disrupted
after a significant contraction of the planetary orbits. 

To illustrate this, the evolution of the semi-major axes  and
eccentricities  of the two planets  for $f_{\mathrm{tid}}=1$ are plotted in Fig. \ref{fig:1}.
In this case  a 2:1 commensurability is first formed which is then disrupted
after the planets have migrated inwards together contracting their orbits
by $\sim 30\%.$ This resonance is disrupted and 
a 3:2 commensurability is then formed.
 The evolution 
 for $f_{\mathrm{tid}}=64$
  is illustrated in  Fig. \ref{fig:64}.
In this case the planets pass through the 2:1 commensurability.
Subsequently the configuration sequentially  forms
 3:2, 4:3 and 5:4 commensurabilities. Each of these are associated
with significant contractions of the orbits.
  The  case with
  $f_{\mathrm{tid}}=1024$
 is illustrated in Fig. \ref{fig:1024}.
 In this case the system passes through the 2:1, 3:2 and 4:3 commensurabilities.
Subsequently 
 5:4, 6:5 and 7:6 commensurabilities are formed each of which
is associated with a significant orbit contraction.
  The  results for
 $f_{\mathrm{tid}}=4096$
   are shown in Fig. \ref{fig:4096}.
 No lasting commensurability is formed until $p=10.$ Again
this commensurability survives while the orbits undergo a large contraction.
This commensurability is of somewhat higher degree than would be expected  from \cite{gl93}.
It is likely that it can survive on account of the eccentricity damping employed here.

  In Table \ref{table1} we also indicate the first order commensurability of lowest degree
that is estimated to be possible by use of  the inequality (\ref{ehcons})
 together with  an  estimate made  assuming the maximum allowed
$\langle \sin^2\phi_i\rangle \le 0.1.$ All the simulation results are  consistent
with the inequality (\ref{ehcons}). If it is used directly, the lowest degree
commensurability that is possible to form is, as expected, somewhat underestimated.
However, if we assume the  maximum allowed
$\langle \sin^2\phi_i\rangle \le 0.1,$ Table \ref{table1}  indicates that  the lowest degree
commensurability that is possible to form is somewhat overestimated 
for the larger values of  $f_{\mathrm{tid}}.$
\section{Conclusion}
We have studied  the development  of orbital resonances
by migrating planets in the terrestrial mass range for  migration time scales varying between
$\sim 10^6 $~yr and $\sim 10^3 $~yr. 
In a typical simulation, a sequence of resonances  occurred, each of which 
could survive significant orbital evolution before being lost.
For the slowest migration rates considered, a 2:1 commensurability was able to form.
At the fastest rates only commensurabilities with large $p$ persisted.
We also found analytic conditions for
first order commensurabilities to be formed.  These were consistent with our
numerical simulations.




\begin{thebibliography}{99}





\bibitem
[{Brouwer \& Clemence} {1961}]{bc61}
Brouwer, D., Clemence, G. M., 1961, {\it Methods of Celestial Mechanics},
Academic Press, New York


\bibitem
[Gladman (1993) ]{gl93}
Gladman, B., 1993 \icar 106, 247 


\bibitem
[{Goldreich \& Tremaine} {1980}]{gt80}
Goldreich, P., Tremaine, S., 1980, \apj 241, 425



\bibitem
[{Kley, Peitz \& Bryden} {2004}]{kpb04}
Kley, W., Peitz, J., Bryden, G. 2004, \ana 414, 735



\bibitem
[{Lee \& Peale} {2002}]{lp02}
Lee, M.H., Peale, S.J., 2002, \apj 567, 596


\bibitem
[{Lin \& Papaloizou} {1986}]{lpap86}
Lin, D.N.C., Papaloizou, J.C.B., 1986, \apj 309, 846

\bibitem
[{ Marcy et~al.} {2001}]{mbfv00}
Marcy, G.W., Butler, R.P., Fischer, D., Vogt, S.S.; Lissauer, J.J., Rivera, E.
J., 2001, \apj 556, 296

\bibitem
[{ Mayor et~al.} {2004}]{munp00}
Mayor, M., Udry, S., Naef, D., Pepe, F., Queloz, D., Santos, N. C., Burnet, M.,
2004, \ana 415, 391

\bibitem
[{ Mayor et~al.} {2008}]{mulp08}
Mayor, M., Udry, S., Lovis, C., Pepe, F., Queloz, D., Benz, W., 
Bertaux, J. -L., Bouchy, F., Mordasini, C., Segransan, D., 2008, arXiv0806.4587

\bibitem
[{ McArthur et~al.} {2004}]{mmfq00}
McArthur, B.E.; Endl, M., Cochran, W.D.,
Benedict, G.F., Fischer, D.A., Marcy, G.W., Butler, R.P., Naef, D., 
Mayor, M., Queloz, D., Udry, S., Harrison, T. E.,
2004, \apj 614, L81



\bibitem
[{Nelson \& Papaloizou} {2002}]{np02}
Nelson, R.P., Papaloizou, J.C.B., 2002, \mnr 333, 26



\bibitem
[{Paardekooper \&Papaloizou} {2008}]{pp08}
Paardekooper, S.-J., Papaloizou, J.C.B.,  2008, \ana  485, 877




\bibitem
[{Papaloizou} {2003}]{pa03} 
Papaloizou J.C.B., 2003, {\it Cel. Mech. and Dynam. Astron.}, 87, 53

\bibitem
[{Papaloizou \& Larwood} {2000}]{pl00}
Papaloizou, J.C.B., Larwood, J.D., 2000, \mnr  315, 823

\bibitem
[{Papaloizou \& Terquem} {2006}]{pt06}
Papaloizou, J.C.B., Terquem, C., 2006, Rep. Prog. Phys., 69, 119 


\bibitem
[{Papaloizou \& Szuszkiewicz} {2005}]{ps05}
Papaloizou, J.C.B., Szuszkiewicz, E., 2005, \mnr  363, 153


\bibitem
[{Sinclair} {1975}]{si75}
Sinclair, A. T., 1975, \mnr 171, 59

\bibitem
[{Snellgrove, Papaloizou \& Nelson} {2001}]{spn01}
Snellgrove, M., Papaloizou, J.C.B., Nelson, R.P., 2001,
\ana  374, 1092




\end{thebibliography}
\end{document}